\documentclass[aps,pre,twocolumn,showpacs,superscriptaddress]{revtex4-1}

\usepackage{graphicx}
\usepackage{color}
\begin{document}


\title{Performance of multifractal detrended fluctuation analysis on short time series }


\author{Juan Luis L\'opez}
\affiliation{Departamento de F\'{\i}õsica Aplicada, 
Centro de Investigaci\'on y de Estudios Avanzados del Instituto Polit\'ecnico Nacional,\\
Unidad M\'erida, A.P. 73 Cordemex, 97310 M\'erida, Yucat\'an, M\'exico}
\author{Jes\'us Guillermo Contreras}
\affiliation{Departamento de F\'{\i}õsica Aplicada, 
Centro de Investigaci\'on y de Estudios Avanzados del Instituto Polit\'ecnico Nacional,\\
Unidad M\'erida, A.P. 73 Cordemex, 97310 M\'erida, Yucat\'an, M\'exico}
\affiliation{Faculty of Nuclear Sciences and Physical Engineering, Czech Technical University in Prague, Prague, Czech Republic}

\date{\today}

\begin{abstract}
The performance of the multifractal detrended analysis on short time series is evaluated for
 synthetic samples of several mono- and multifractal 
models.  The reconstruction of the generalized Hurst exponents
is used to determine the range of applicability of the method and the precision of its results
as a function of the decreasing length of the series. As an application
 the series of the  daily exchange rate between the US Dollar and the Euro is studied.
\end{abstract}


\maketitle

\section{Introduction}
\label{sec:intro}

There are many processes of interest in nature and in society which exhibit a fractal or multifractal behavior 
\cite{Mandelbrot79,Mandelbrot82,Grass83,Stan88,BundeHavlin,Mandelbrot97,Stan99}. 
One source of information from these processes are time series obtained from  records of measurements or observations.
These time series may be affected from experimental or observational non-stationary uncertainties which have to be disentangled 
from the potential intrinsic fluctuations and correlations of the studied system. This is a very complex task and many methods
to achieve this goal have been proposed  \cite{Kantelhardt2009}. 

One method which has proved to be quite useful to
detect reliably  long-range correlations in data with trends is the detrended fluctuation analysis (DFA) introduced by Peng et al. 
\cite{DFA}. 

Later, this method has been generalized
to the analysis of multifractal time series  (MFDFA) by Kantelhardt
et al. \cite{MFDFA}  and has been extended to multi-dimensional series \cite{MFDFAhd} and to investigate the power-law correlations between simultaneously-recorded time series  \cite{DXA,Pod09,MFDXA}. MFDFA has been compared favorably to other methods 
\cite{MFDFAComp,MFDFAriver}  and applied to a wide range of fields. Just to name a few cases -- currently \cite{MFDFA} has been cited hundreds of times --
MFDFA has been used to study series from geophysics \cite{MFDFAriver,MFDFAriver2,Kavasseri2005165}, physiology \cite{Makowiec2006632,Dutta,Makowiec11}, financal markets
 \cite{MFDFAstock,MFDFAstock2}, and, of particular interest here,
  to study the exchange rate of different currencies \cite{MFDFAUSDRial, MFDFAUSD,MFDFAasia}.

MFDFA works very well for
time series with some $2^{16}$ elements or more, but nevertheless it is important to evaluate the performance of this, and any other method (see e.g. \cite{lapre79})
 on shorter time series, mainly for two reasons: first,  there are many records  of interest which
are short and second, there are processes for which long records are available, but where it is expected that the multifractal 
behavior changes with time and the study of short fragments of these long series could yield important insight on those
cases.

In this work the performance of MFDFA is studied as a function of the decreasing  length of the series.
The evaluation is performed using computer simulated data sets with known fractal and multifractal behavior.
The results are applied to the analysis of the daily exchange rate between the US Dollar and the Euro. This time series is relatively short, around 3500 entries, given that the Euro currency debuted at
the beginning of 1999. Furthermore in its short life the Euro has gone through a dubitative start,
followed by a strong couple of years and since around 2008 it has been immersed in
a crisis which has threatened its existence.    This turbulent history makes it interesting
to ask if its dynamics have changed with time.

The paper is organized as follows: in the next section the MFDFA method is briefly described
and the notation used in the rest of the paper is introduced. Section \ref{s:Models} presents
the analysis of the mono- and multifractal synthetic data. Section \ref{s:Application}
discusses the application of the results to a time series from finance, namely the daily
exchange rate between the US Dollar and the Euro. Finally, the conclusions of this work
are presented in Section \ref{s:Conclusions}.

\section{Multifractal detrended functional analysis}
\label{sec:mfdfa}

The MFDFA method introduced in \cite{MFDFA} will be described briefly here. The input to the method is 
a time series $x(i)$ of finite length $N$.
It is assumed that the time series has a compact support; i.e., that only a negligible fraction of the elements $x(i)$ 
are zero. The algorithm has 5 steps:
\begin{enumerate}
\item Compute the profile $Y(j)$, where $j = 1$,...,$N$:
\begin{equation}
Y(j) = \sum^j_{i=0} \left[ x(i)-<x> \right]. 
\end{equation}
\item Divide the new series $Y(j)$ in $N_s$ non-overlapping contiguous segments of size $s$ starting from the beginning of the series and then repeat starting from the end to obtain $2N_s$ segments.
\item Calculate, for all segments $\nu$ and all sizes $s$, the 
local polynomial trend of order $m$, $P^m_\nu$, via a least-square fit and compute the variance:
\begin{equation}
F^2(\nu,s)=\frac{1}{s}\sum^s_{i=1}\left\{ Y\left[(\nu-1)s+i\right] - P^m_\nu(i)\right\}^2.
\end{equation}
\item Average over all segments of a given size $s$ to obtain the $q$-order fluctuations:
\begin{equation}
F_q(s) = \left\{ \frac{1}{2N_s}\sum^{2N_s}_{\nu=1}\left[ F^2(\nu,s)\right]^{q/2}\right\}^{1/q},
\end{equation}
or, for $q=0$,
\begin{equation}
F_0(s)=\exp\left\{\frac{1}{4N_s}\sum^{2N_s}_{\nu=1}\ln\left[ F^2(\nu,s)\right] \right\}.
\end{equation}
\item For signals with fractal properties there is  a range of sizes, $s_{\rm min} <s < s_{\rm max}$, at a given order $q$ for which
\begin{equation}
F_q(s) \sim s^{h(q)}.
\label{eq:defhq}
\end{equation}
The $h(q)$ are called {\it generalized Hurst exponents} and are the output of the MFDFA algorithm. Note that $h(q)$ is related to the  singularity spectrum $f(\alpha)$, where $\alpha$ is called the H\"older exponent, through the following relations:
\begin{equation}
\alpha=h(q)+qh^\prime(q),
\end{equation}
and
\begin{equation}
f(\alpha)=q\left[\alpha-h(q)\right]+1,
\end{equation}
where $h^\prime(q)$ denotes the derivative of $h$ with respect to $q$.

\end{enumerate}

\section{Performance on synthetic data\label{s:Models}}

In this section synthetic signals are used to evaluate the performance of the MFDFA method as a function of length. The goal of this section is to get an insight on what is the shortest length of  series from each model that can be reliably analyzed; what is the magnitude of the precision that can
be expected for such a length and in which range of $q$ is the analysis valid. Both  mono- and multifractal models are studied and compared to the corresponding analytic predictions. 

Note that these studies yield only estimations of possible shortest lengths and precision of the analysis and not  definite predictions, because real time series are much more complex than
their synthetic counterparts. On the other hand, these studies show where it is necessary to be
specially careful when assigning a mono- or multifractal behavior to a real time series or
when assessing the amount of multifractality present in a real time series.

Note that in all the synthetic cases studied here 
the signals were detrended with a polynomial of order two in the third step of the MFDFA method explained 
in section \ref{sec:mfdfa}.

\begin{figure}[b!]
\begin{center}$
\begin{array}{c}
\includegraphics[width=0.5\textwidth]{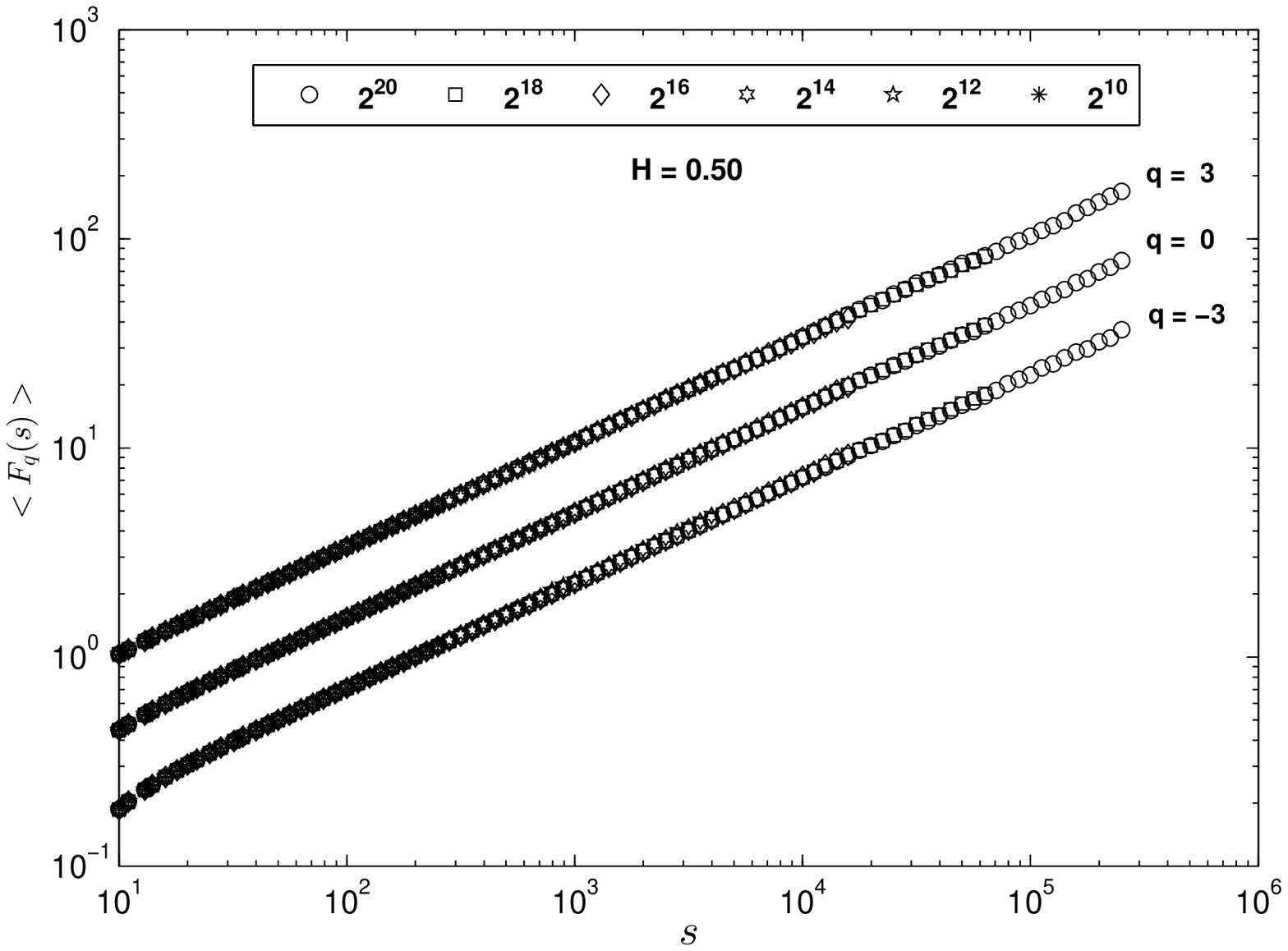} \\
\includegraphics[width=0.5\textwidth]{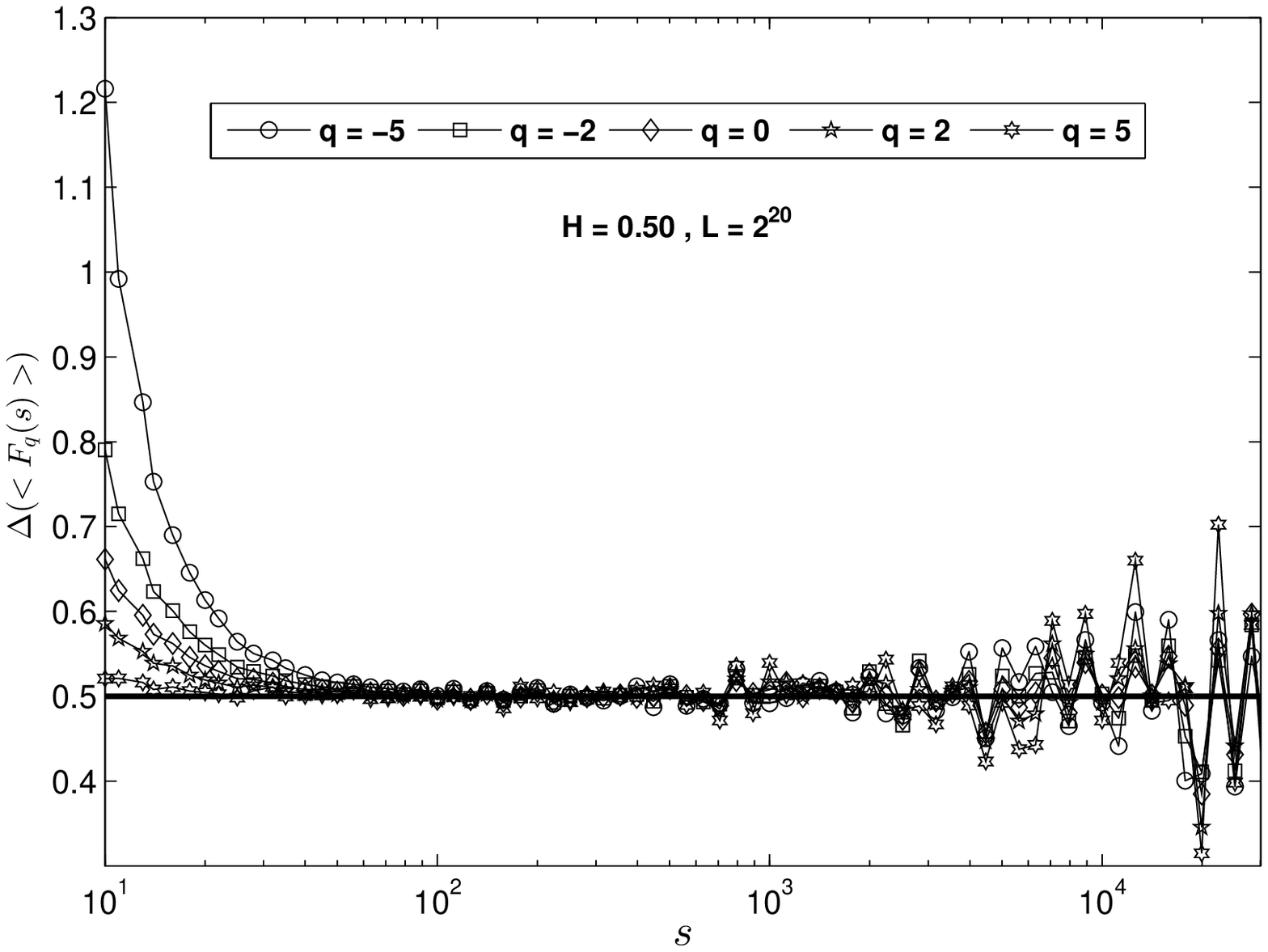} 
\end{array} $
\end{center}
\begin{picture}(0,0)
   \put(-115,370){\bfseries (a)}
   \put(-115,185){\bfseries (b)}
\end{picture}
\caption{The case of white noise: {\bf (a)} Representative average fluctuations of order $q$ as
a function of $s$ for 
series of six different lengths. (For clarity the lines for $q=0$ and $q=3$ were multiplied by a factor of 2 and 4 respectively.) {\bf (b)} Local derivative of the average fluctuations 
as
a function of $s$ for series of
length $2^{20}$ and different values of $q$.}
\label{fig:H050comp}
\end{figure}

\subsection{Computer generated time series}

Series of length $2^k$ with $k=$ 20, 18, 16, 14, 12 and 10
were generated
for each one of  the synthetic models that were analyzed. In each case the number of independent realizations were 10, 40 and 100 
for $k=$ 20, 18 and 16--10 respectively.

Monofractal models are interesting to evaluate the performance of a method because they have the simplest
functional form for the generalized Hurst exponents: a constant. Three different models were studied. The case
of white noise characterized by a Hurst exponent $H=0.5$ and
the absence of long range correlations; and
the cases   $H=0.75$ and $H=0.25$ which present long-range correlations and anti-correlations respectively.

Multifractal models exhibit richer behavior in the generalized Hurst exponent presenting thus a harder
challenge to the MFDFA method.  Three types of stochastic 
binomial cascades were used as multifractal models: log-Poisson, log-Gamma and log-Normal. All three have
been shown to have different and interesting multifractal behavior as discussed for example in \cite{Mandelbrot89}.

\begin{figure}
\begin{center}$
\begin{array}{c}
\includegraphics[width=0.5\textwidth]{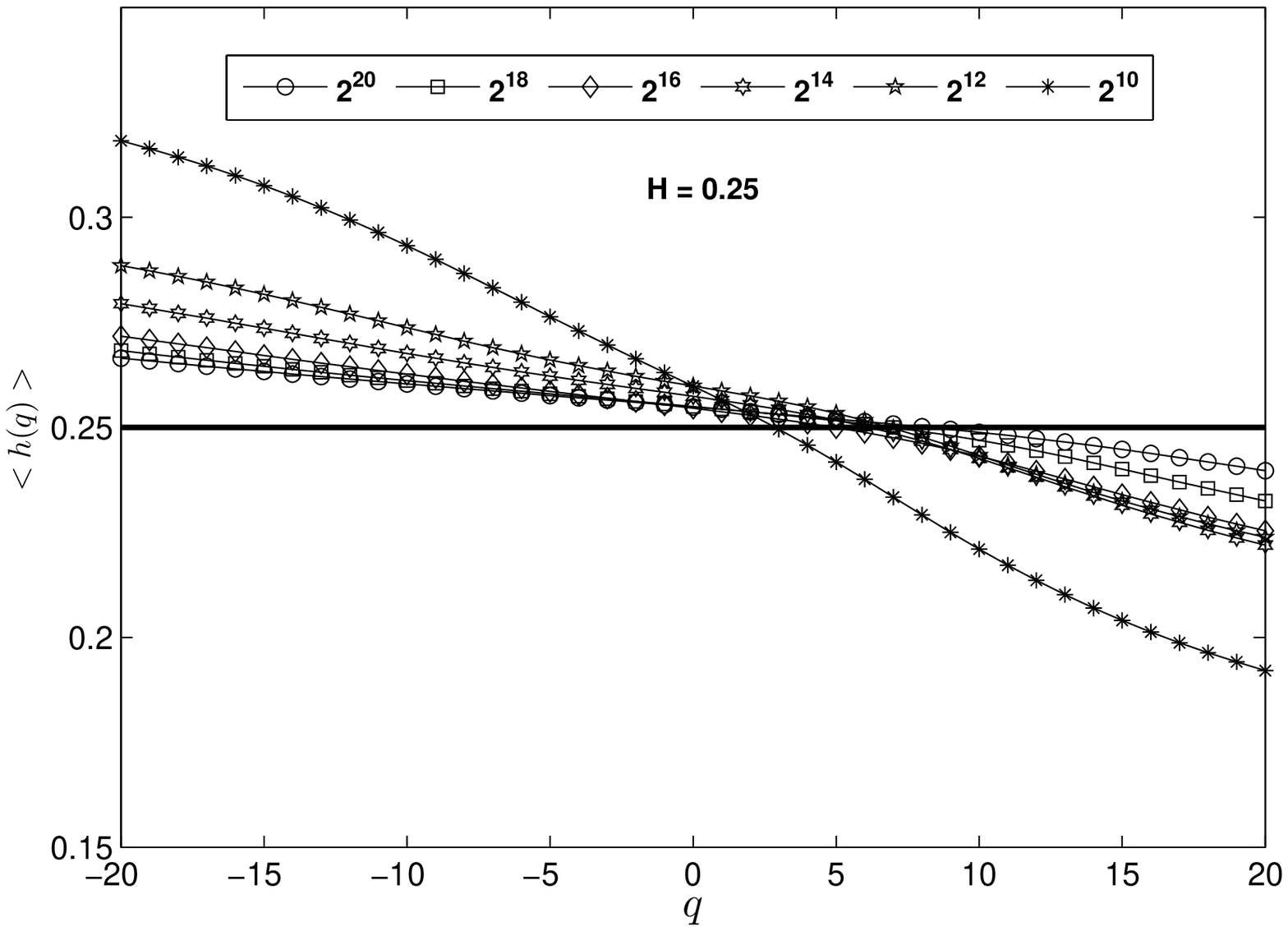}\\
\includegraphics[width=0.5\textwidth]{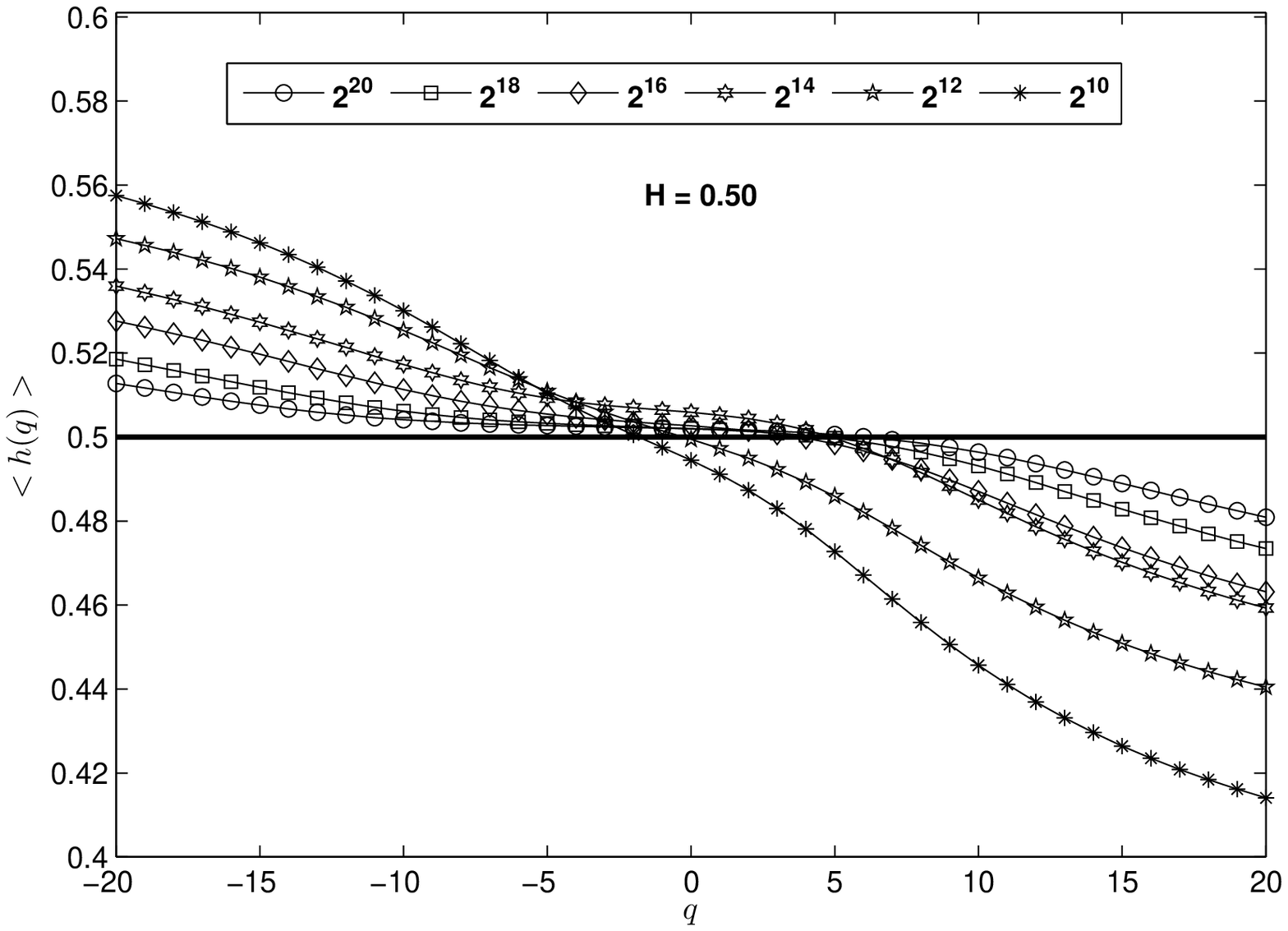} \\
\includegraphics[width=0.5\textwidth]{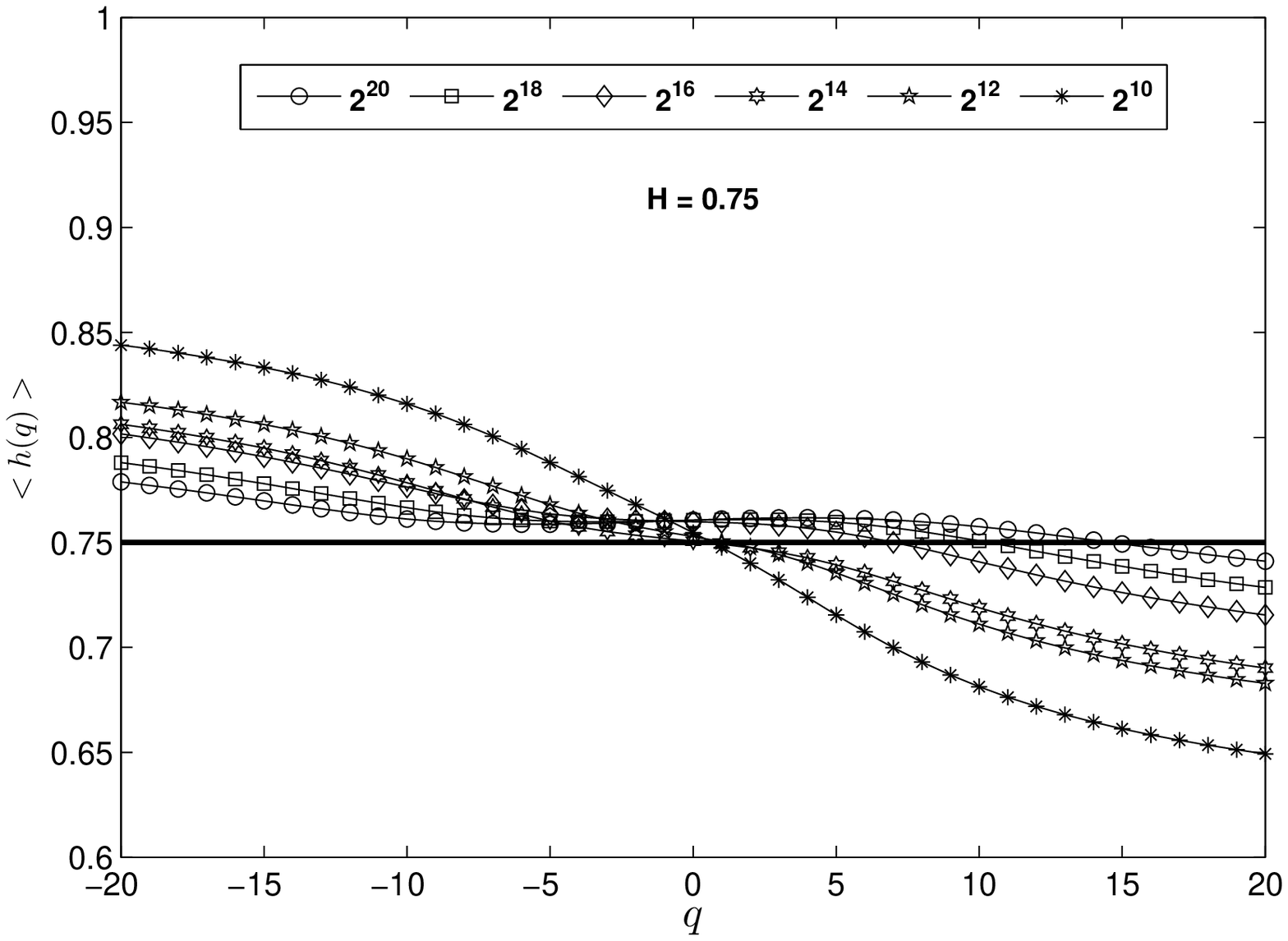} \\
\end{array} $
\end{center}
\begin{picture}(0,0)
   \put(-125,550){\bfseries (a)}
   \put(-125,370){\bfseries (b)}
   \put(-125,185){\bfseries (c)}
\end{picture}
\caption{
Mean generalized Hurst exponent for {\bf (a)} $H=0.25$, {\bf (b)} $H=0.50$ and {\bf (c)} $H=0.75$
and for different lengths of the series compared to the theoretical expectation represented by the solid line.}
\label{fig:hqHurst}
\end{figure}

\subsection{Monofractal signals}

The key assumption of the MFDFA method is that $F_q(s) \sim s^{h(q)}$ for some range in $s$, so that $h(q)$ can
be extracted, in that $s$ range,
 by a fit to a line in a log-log scale. This has been shown to be the case for long mono-fractal series
in a wide range of $q$; e.g., \cite{MFDFA,MFDFAComp}. Here the main interest is the dependance on the decreasing
length of the series. In particular it is important to determine if the range in $s$ depends on the length of the series.
To obtain a statistically stable answer to this question, the average $F_q(s)$, denoted by 
$\left<F_q(s)\right>$, of all  independent realizations of
a given model was used.

The behavior for the case of white noise, $H=0.5$, is depicted in Figure \ref{fig:H050comp}.
The upper panel in Figure \ref{fig:H050comp} shows  $\left<F_q(s)\right>$ for three  representative values of
$q$ and the six lengths studied in this work.
There are two important observations to be made from this panel: ($i$) there is a range of box sizes $s$ where $F_q(s)$ exhibits a power
law behavior and, ($ii$) the fact that the symbols for the different lengths can not be
distinguish by eye  means that this power law behavior does not depend on the length of the series, at least for the lengths considered here. 

Remember that the fifth point of the MFDFA algorithm requires the range of sizes where the
power law behavior is valid in order to extract from this range the generalized Hurst exponent.
Looking again at the upper panel in Figure \ref{fig:H050comp} it is observed that, as 
expected, the choice of $s_{\rm max}$ depends on the length of the series, while 
the choice of $s_{\rm min}$ depends on $q$.

To make easier to visualize this last point, the lower panel in Figure \ref{fig:H050comp} shows 
the local difference $\Delta(\left<F_q(s)\right>)$ defined as 
\begin{equation}
\Delta(\left<F_q(s)\right>) = \frac{\ln(\left<F_q(s_{i+1})\right>) -
\ln(\left<F_q(s_{i})\right>)}{\ln(s_{i+1})-\ln(s_i)}
\end{equation}
where $i$ runs over all sizes. $\Delta(\left<F_q(s)\right>)$ is shown
for different values of $q$ at a fix length
 $L=10^{20}$. 
It is clear that $\Delta(\left<F_q(s)\right>)$ is constant over a large range of sizes $s$.
At larger values of $s$ there are strong fluctuations due to the small number of boxes that can
be formed at large sizes.  At lower values of $s$
the behavior of $\Delta(\left<F_q(s)\right>)$ depends on $q$. Note that a
 qualitatively similar behavior is found for the monofractal models defined by $H=0.25$ and $H=0.75$.
 
To simplify the analysis a value of $s_{\rm min}$ was chosen so that  the 
dependance on $q$ was avoided. The
actual values used on the fits were: $s_{\rm min} = 50$, 40 and 30 for $H=0.25$, 0.50 and 0.75 respectively. The
values of $s_{\rm max}$ used to extract $h(q)$ were $s_{\rm max}=10^4$ for lengths $2^k$ with $k=16$, 18 and 20; $s_{\rm max}=3000$ for $k=14$;
$s_{\rm max}=800$ for $k=12$ and $s_{\rm max}=200$ for $k=10$. These values were used for all three Hurst parameters. 


Within the stipulated $s$ regions each realization of a series presented the power like behavior of equation (\ref{eq:defhq}) and for each realization
the values of the generalized Hurst exponents were extracted by a linear least squares fit. These values were averaged over all realizations
of a given value of $H$ and a given length. The average, denoted $\left<h(q)\right>$,  was compared to the theoretical expectations. The results for all three cases are shown in Figure \ref{fig:hqHurst}. The general trend is the same for the
three cases: ($i$) the values of $h(q)$ are under/over predicted  at large/small values of $q$ and the best agreement
are for values of $q$ close to zero,  ($ii$) the concordance between theory and simulation is best for long series and
deteriorates as the length of the series decreases.

The first trend mentioned implies that in case large ranges in $q$ are studied, the results
would mimic a multifractal behavior specially if the amount of multifractality is estimated from
the difference in the values of $h$ at large negative  and positive values of $q$ as sometimes
is done. 

If the analysis
is restricted to a more central region in $q$, say $|q|<5$, then MFDFA yields good results
for all lengths. In this range of $q$ the agreement
of theory and simulation for $H=0.25$ is between 2\% at $q=5$ and 8\% at $q=-5$ for
series of $2^{12}$ elements and it goes up to 10\% at $q=-5$ for the shortest series while
is between 1\% and 3\% for the longest series. For the case $H=0.5$ the worst agreement
happens 
for the shortest series and it is just 5\%, but in general the results are within 3\% of the
theoretical expectations. The situation is even slightly better for $H=0.75$ with most
comparisons between $h(q)$ as predicted by theory and found with the simulations
below 2\% and the worst cases, for the shortest series at $q=5$ and $q=-5$ only 6\% away
of the predictions.

The previous discussion referred to the average $h(q)$. The fluctuation between the different
values of $h(q)$ corresponding to each realization have been evaluated with the standard deviation. For lengths of $2^{14}$ or longer the standard deviation is well below 5\% over all
values of $q$. For the shorter series and $H=0.25$
 the fluctuations for $-5<q<5$ reach 10\% and 20\%
for lengths $2^{12}$ and $2^{10}$ respectively. The situation is better for $H=0.5$ and $H=0.75$
where the corresponding fluctuations are 8\% and 15\%; and 6\% and 12\% respectively.
\begin{figure}
\begin{center}$
\begin{array}{c}
\includegraphics[width=0.5\textwidth]{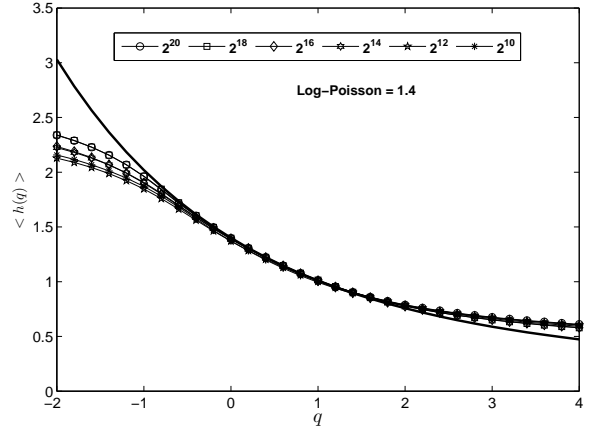} 
\end{array} $
\end{center}
\caption{
Mean generalized Hurst exponent for Log-Poisson binomial cascades with parameter 1.4
for different lengths of the series compared to the theoretical expectation represented by the solid line.}
\label{fig:hqPoisson}
\end{figure}

\subsection{Multifractal signals}

Following \cite{Mandelbrot89}, stochastic binomial cascades were used as models of multifractal behavior. The cascades are built as follows.
Consider a unit of some property, commonly named {\em mass}, in the interval $[0,1]$. Next split
the interval in two halves  and assign a random fraction of the mass to each of the new intervals. Repeat
the procedure for each half. After $k$ steps the mass is distributed in $2^k$ intervals of size $2^{-k}$ yielding a series of length $2^k$. 
The assignment of the random fraction to each half is not arbitrary, the density function providing
the random fractions has to conserve the mass in the average.
The random fractions are independent,
identically distributed random variables drawn from a
specific distribution. Mandelbrot  has 
shown (see \cite{Mandelbrot89} and references therin) that this process produces signals with multifractal
properties.

 Analytical results for the multifractal
properties of stochastic binomial cascades are available for a number of distributions  \cite{Mandelbrot89}.
Here the cases of Log-Normal, Log-Gamma and Log-Poisson cascades are used to evaluate
the performance of the MFDFA method on short time series. Long time series for
three of the five  models analyzed here have been studied in \cite{MFDFAComp} and those
results agree with the findings below.

In all multifractal cases studied here, a region where equation (\ref{eq:defhq}) was fulfilled could be identified.
The same procedure outlined above was used to obtain the values of $s_{\rm min}$ and $s_{\rm max}$. In the case of the multifractal models the value of $s_{\rm min}$ did not depend on $q$, nor in the length of the series, and a value $s_{\rm min} = 40$ was used in all cases. For $s_{\rm max}$ the same values as for the monofractal signals were used; namely
$s_{\rm max}=10^4$ for lengths $2^k$ with $k=16$, 18 and 20; $s_{\rm max}=3000$ for $k=14$;
$s_{\rm max}=800$ for $k=12$ and $s_{\rm max}=200$ for $k=10$. 

{\bf Log-Poisson cascade}. This  is a random discrete model which depends on only one parameter which represents the
mean and variance of a Poisson distribution. The series studied here were generated with the value 1.4 which ensures that the mass is conserved in the average for binomial cascades. 

The
results of MFDFA for this model are shown in Figure \ref{fig:hqPoisson}. Series of all lengths under investigation reproduced the theoretical prediction with the same precision of better than
half a percent for
$-0.5<q<2.0$. For larger values of $q$ the simulation over estimates the prediction more and more; at $q=4$ the difference between prediction and simulation is 3\% independent of the
lenght of the series.

 For smaller values of $q$ the method is not able to yield the predicted shape of the generalized Hurst exponent and seems to saturate to a value depending on the length of the series. At $q=-1$ the agreement is still of the order of 1\% for the shortest series and better
for the longest one, but at $q=-2$ the difference between prediction and simulation is already
up to 20\% for the longest series and 30\% for the shortest one. 

Note that in this case the estimation of multifractality as the difference of the generalized Hurst 
exponent at a large negative $q$ and a large positive $q$ would yield a smaller multifractality
than expected from  theory . This behavior is the opposite to the behavior shown
by monofractal series which tend to yield a bigger multifractality than what is present in the model.

For this multifractal model, the fluctuations among independent realizations, as quantified
by the standard deviation, are of the order of 6--8\% for lenghts down to $2^{12}$ at large
values of $q$; at $q=0$ the standard deviation grows from 1\% to 8\% for lengths decreasing
from $2^{20}$ to $2^{12}$ and are around 10--15\% for $q=-2$. For the shortest length of
$2^{10}$ the fluctuations are bigger: 15\% at $q=0$, 25\% at $q=4$ and up to 30\% at
$q=-2$

\begin{figure}[b!]
\begin{center}$
\begin{array}{c}
\includegraphics[width=0.5\textwidth]{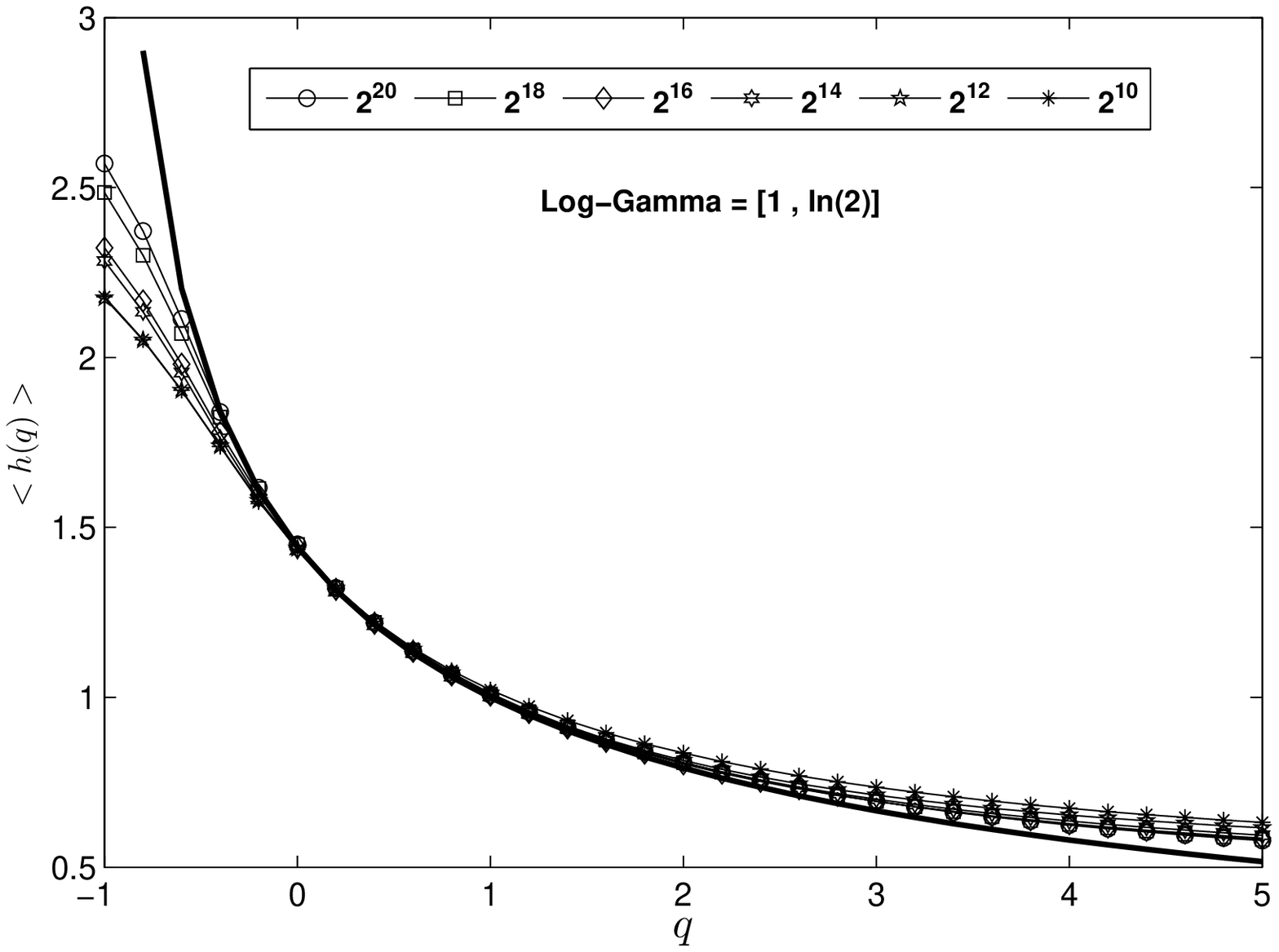}\\
\includegraphics[width=0.5\textwidth]{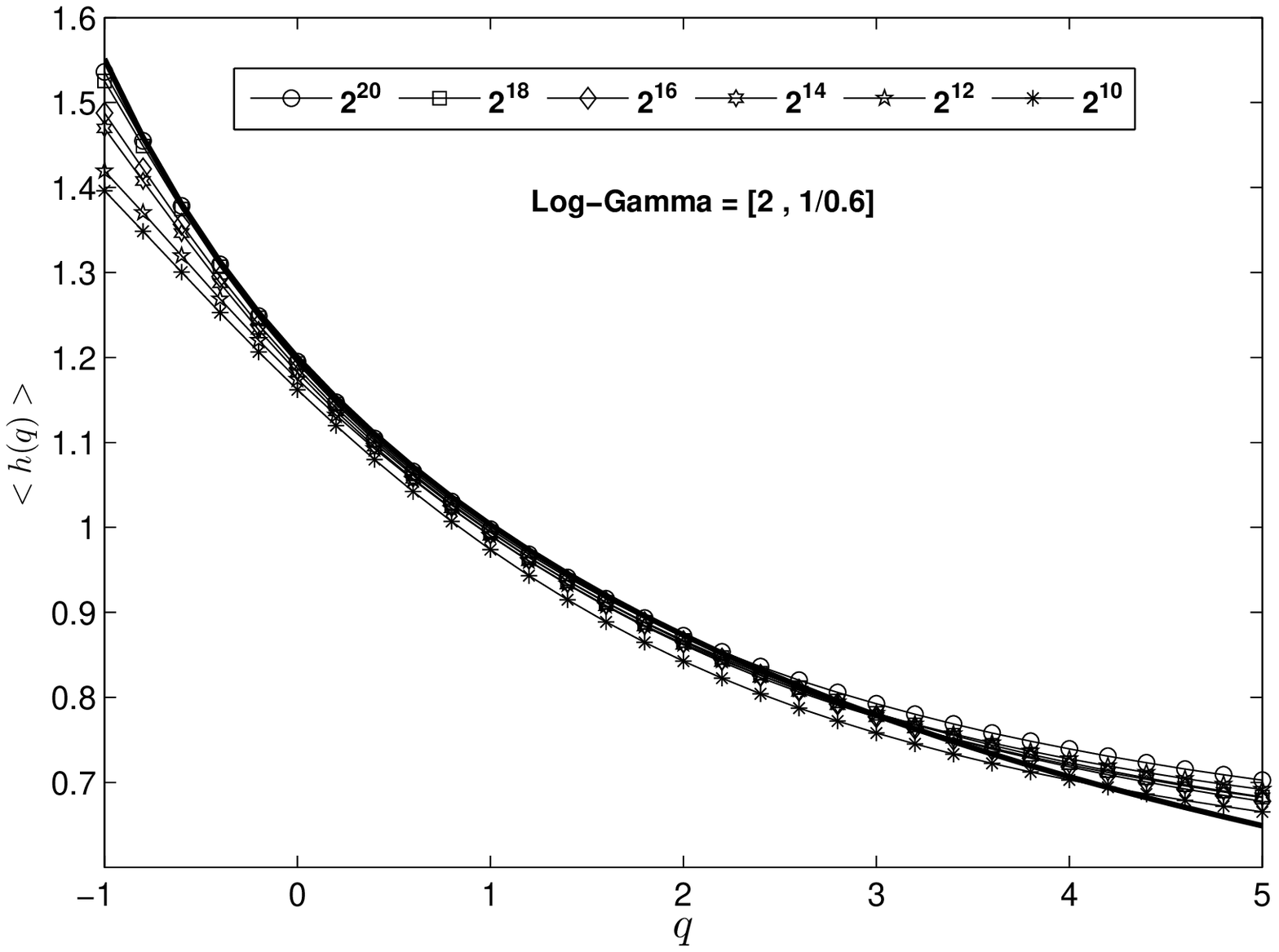} \\
\end{array} $
\end{center}
\begin{picture}(0,0)
   \put(-125,360){\bfseries (a)}
   \put(-125,175){\bfseries (b)}
\end{picture}
\caption{
Mean generalized Hurst exponent for Log-Gamma binomial cascades with parameters {\bf (a)} [1, ln(2)] and {\bf (b)} [2, 1/0.6]
for different lengths of the series compared to the theoretical expectation represented by the solid line.}
\label{fig:hqGamma}
\end{figure}

\begin{figure}[t!]
\begin{center}$
\begin{array}{cc}
\includegraphics[width=0.5\textwidth]{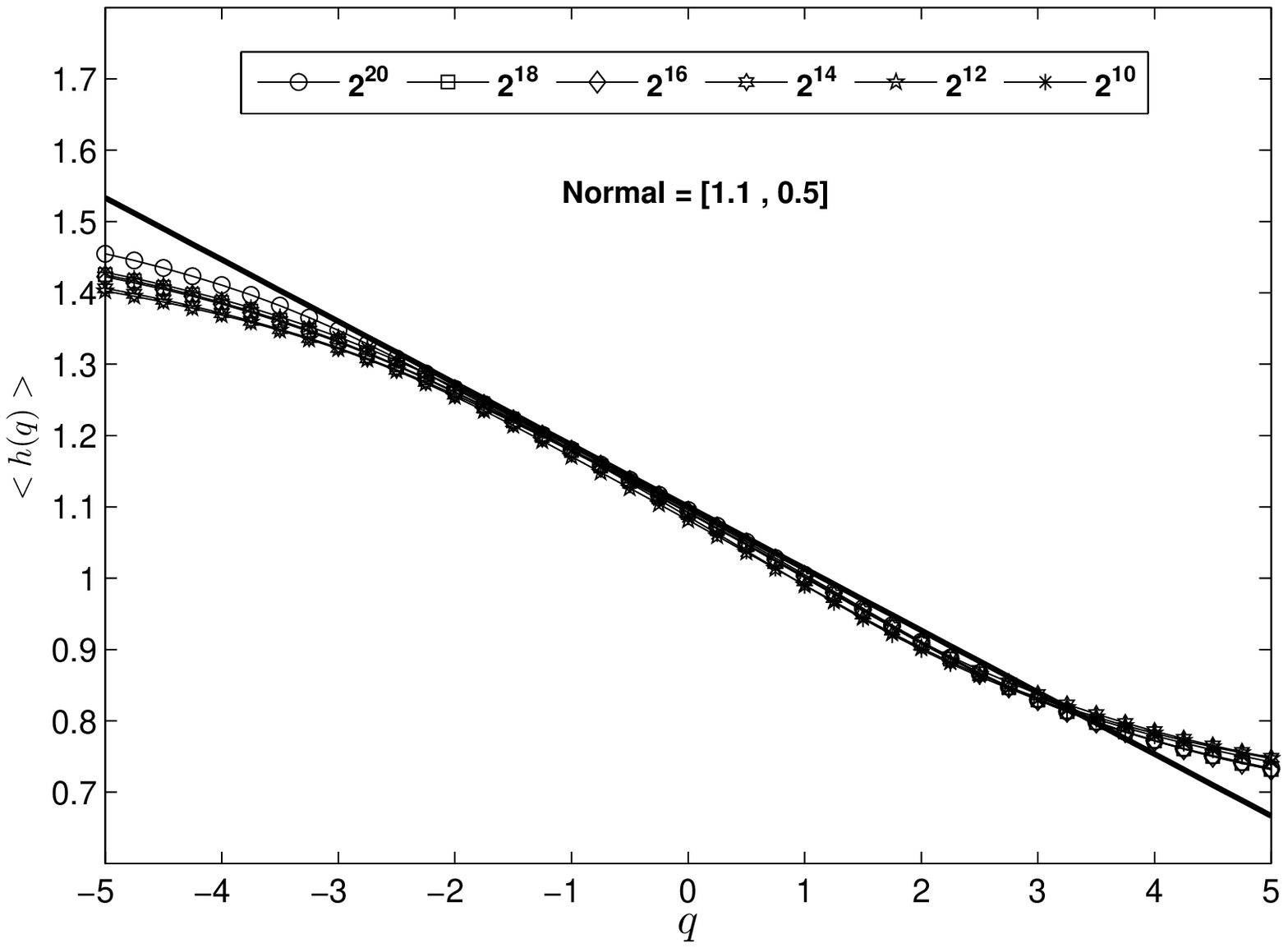}\\
\includegraphics[width=0.5\textwidth]{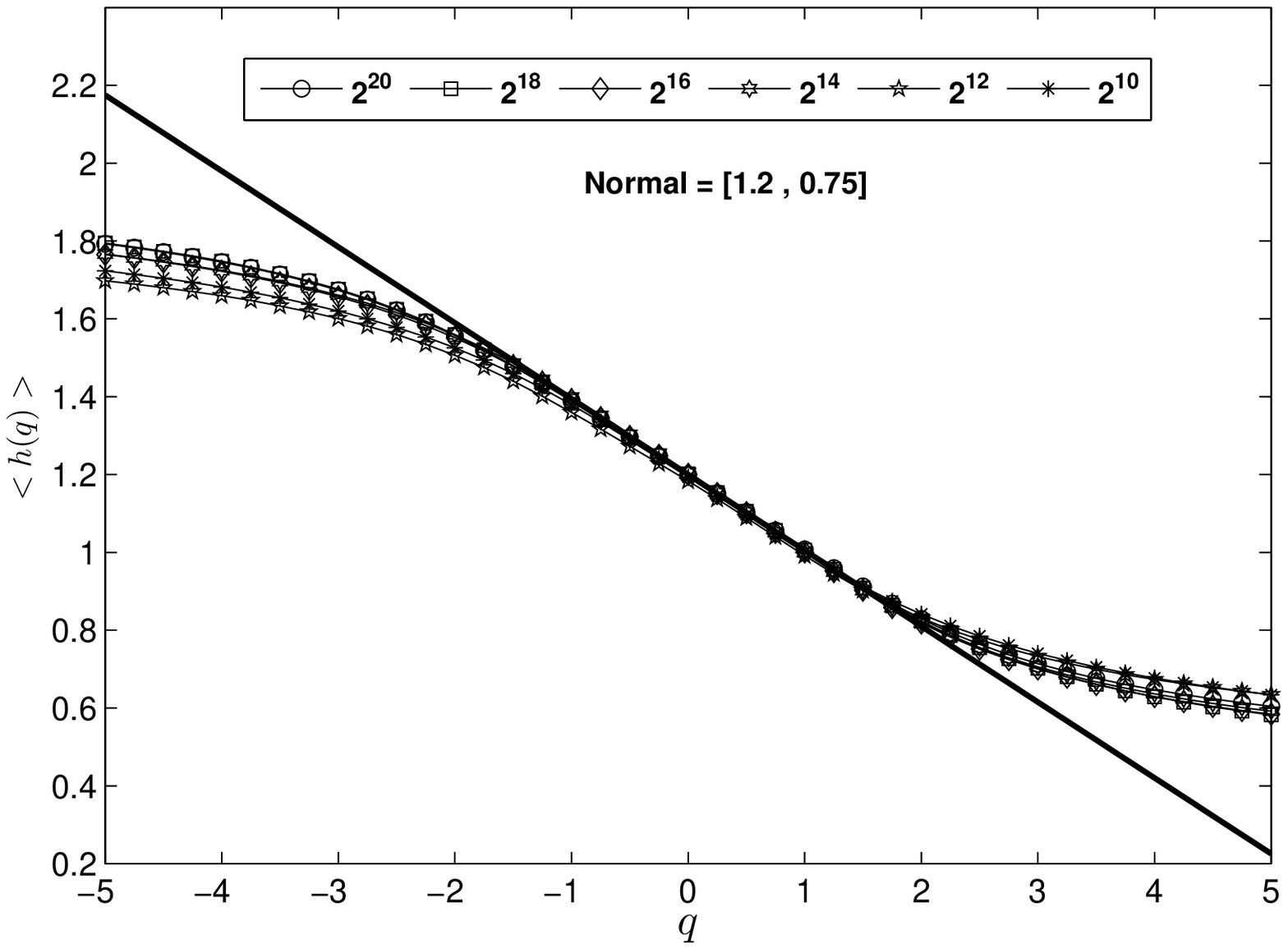} \\
\end{array} $
\end{center}
\begin{picture}(0,0)
   \put(-125,360){\bfseries (a)}
   \put(-125,175){\bfseries (b)}
\end{picture}
\caption{
Mean generalized Hurst exponent for Log-Normal binomial cascades with parameters {\bf (a) }[1.1, 0.5] and {\bf (b)} [1.2, 0.75]
for different lengths of the series compared to the theoretical expectation represented by the solid line.}
\label{fig:hqNormal}
\end{figure}

{\bf Log-Gamma cascade}. This is a random continuum model where the random numbers are taken from a Gamma distribution which is characterized by  the shape and the inverse scale  parameters. Two different sets of parameters were used to evaluate the performance of MFDFA:
[1, ln(2)] and [2, 1/0.6]. 

The results are shown in Figure \ref{fig:hqGamma}. The conclusions are similar to those of the analysis of Log-Poisson cascades: There is good agreement between theory and simulation for all lengths of the series in the middle region of $q$. For large value of
 $q$ the simulation over-estimates the theoretical result while at low values of  $q$ the simulations
 under estimate the prediction. 
 
 The range in $q$ where the agreement between simulation and theory is good depends on the
 parameters. For the parameter set [1, ln(2)] the difference between simulation and theory is 4\%
 for the shortest series and 3\% for the longest series. At $q=-1$ the model diverges, but the
 simulation yields values from 2.2 to 2.6 depending on the length of the series. 
 
 For the second
 set of parameters,  [2, 1/0.6], the agreement is good for all lengths in the range from $q=-1$ to $q=5$. In all
 cases the agreement is below 2\%. 
 
 
 For this multifractal model, the fluctuations among independent realizations, as quantified
by the standard deviation, have a similar behavior for both sets of parameters. 
For the parameter set [1, ln(2)] at $q=0$ they
grow from 1\% to 12\% for lengths from $2^{20}$ to $2^{10}$. For $q=5$ they are around 8-10\%
except for $2^{10}$ which reaches 25\%. For $q\to -1$ the fluctuations are bigger: around 15\%
for all lengths except the shortest where the standard deviation grows to 30\%.
For the second parameter set,  [2, 1/0.6], the qualitative behavior is similar but quantitatively the fluctuations are substantially smaller, being about half of the fluctuations for the parameter
 set [1, ln(2)].

{\bf Log-Normal cascade}. This model is characterized by two parameters which correspond to 
the mean and standard deviation of a Normal distribution.  Two different sets of parameters were used to evaluate the performance of MFDFA:
[1.1, 0.5] and [1.2, 0.75]. The interpretation of these parameters for $h(q)$ is straightforward:
in this case $h(q)$ is a straight line,
the value of $h(q=0)$ corresponds to the mean of the Normal distribution while the slope is
directly related to its variance.

 The results are shown in Figure \ref{fig:hqNormal}. As in the previous multifractal models there is no strong dependance with the length of the series down to the shortest series that were studied. Simulation and theory agree in the
 middle region of $q$ and the range in $q$ where the agreement is good depends on 
 the value of the parameters. At large values of $q$ the simulation yield values larger than expected from the theory and at small values of $q$ the theory is above the simulations. So, also
 in
 these cases  the amount of multifractlity could be under estimated. 
 
 For the first set of parameters, [1.1, 0.5] the agreement between theory and simulation
 in the range from $q=-2$ to $q=3$ is 2\% or better. The range in $q$ with a similar
 agreement between theory and simulation is reduced to the range from $q=-1$ to $q=2$ for the
 second set of parameters, [1.2, 0.75].
  
   For this multifractal model, the fluctuations among independent realizations, as quantified
by the standard deviation, have a similar behavior for both sets of parameters. 
For $|q|=-3$ they grow from 2--3\% to 10\% for lengths decreasing from $2^{20}$ to $2^{12}$
and are  smaller for $q=0$. For the shortest length the fluctuations reach 20\% at
large $|q|$ and around 12\% at $q=0$.

 \begin{figure}[t!]
\begin{center}$
\begin{array}{c}
\includegraphics[width=0.5\textwidth,height=0.25\textheight]{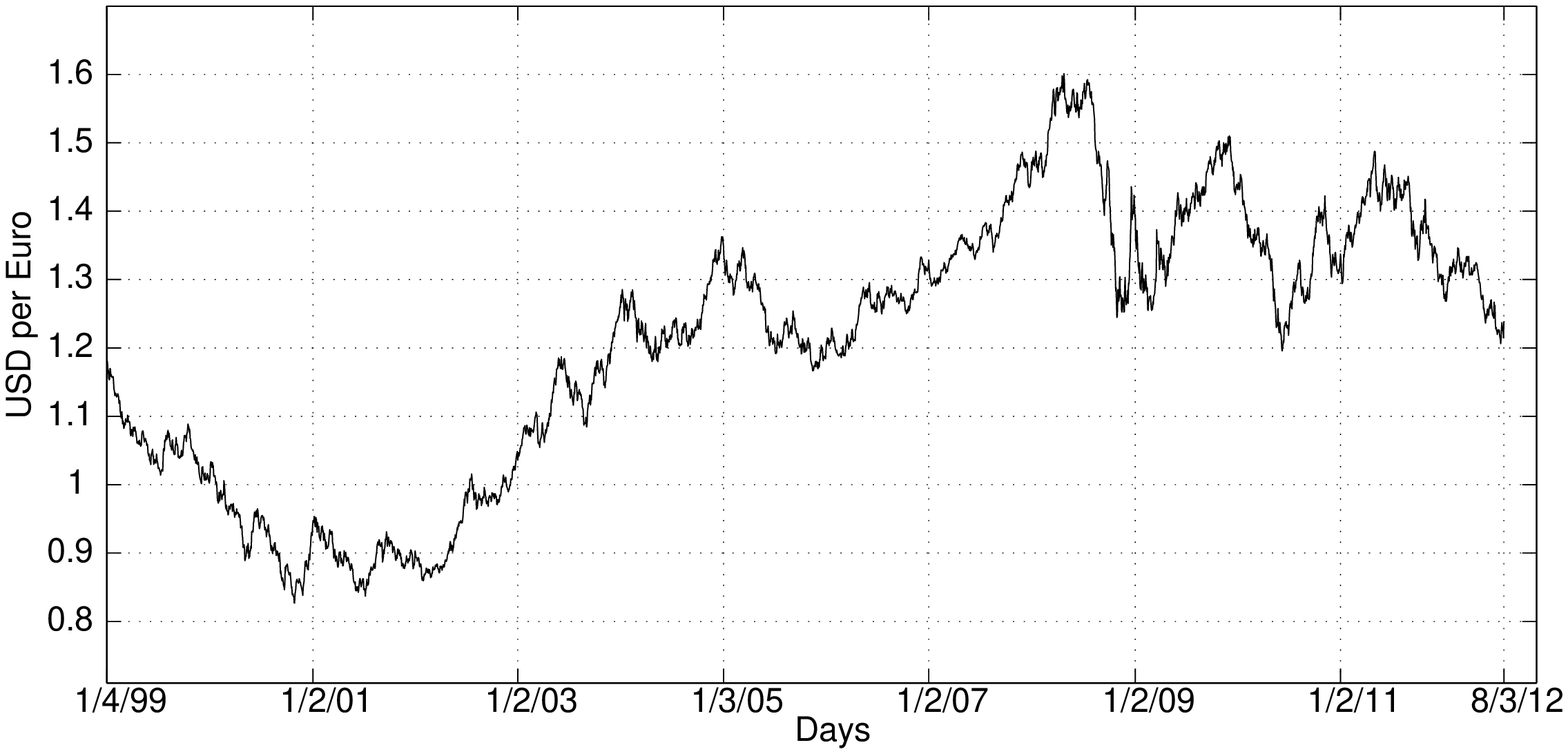}\\
\includegraphics[width=0.5\textwidth,height=0.25\textheight]{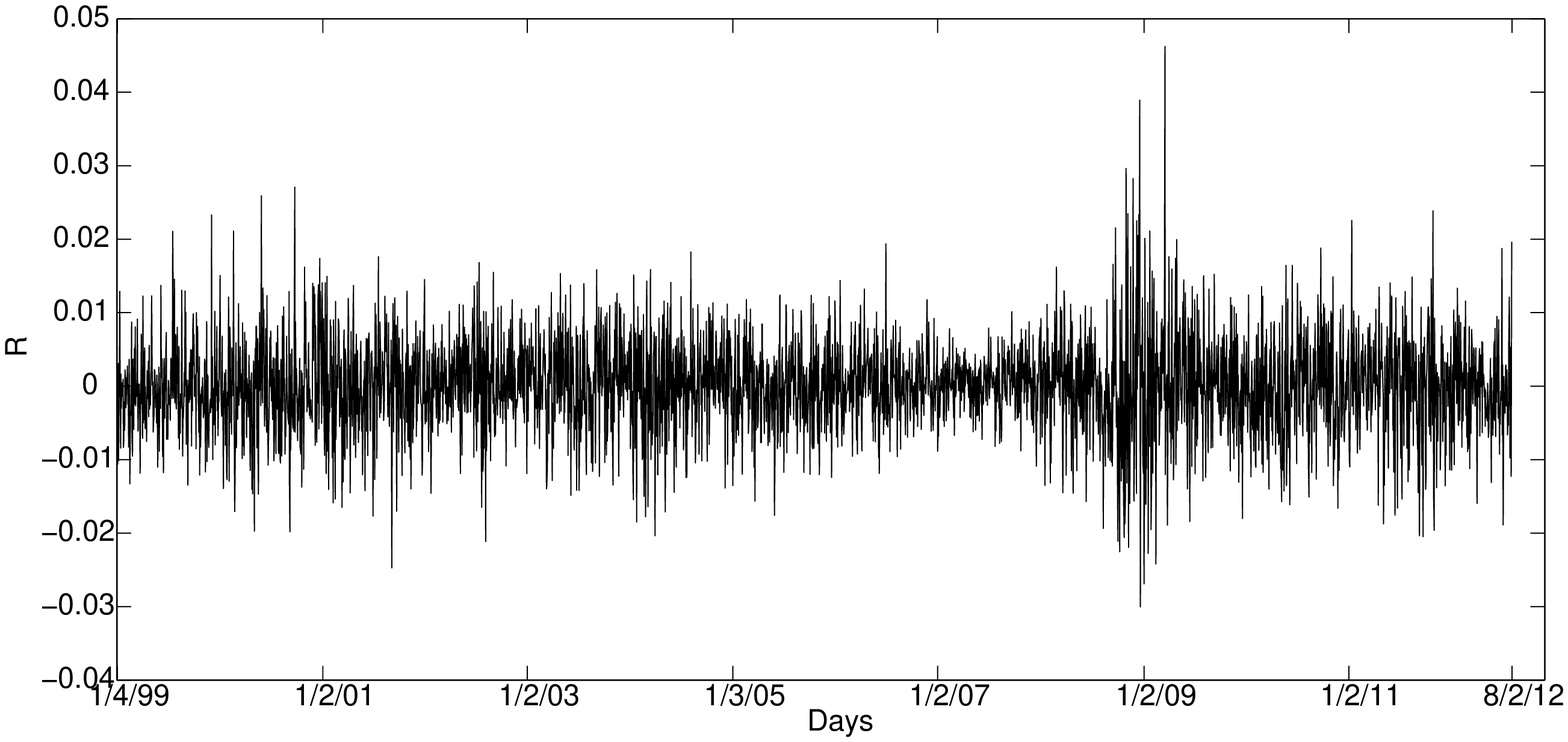} \\
\end{array} $
\end{center}
\begin{picture}(0,0)
   \put(-125,340){\bfseries (a)}
   \put(-125,165){\bfseries (b)}
\end{picture}
\caption{
Exchange rate between the US Dollar and the Euro. ({\bf a}) The original time signal
containing 3420 data spanning from January 4th, 1999 to August 3rd, 2012. ({\bf b})
The time series of the logarithmic differences of consecutive exchange rates used as input to
MFDFA.}
\label{fig:hqEURUSD}
\end{figure}

 \begin{figure}[t!]
\begin{center}$
\begin{array}{c}
\includegraphics[width=0.5\textwidth,height=0.25\textheight]{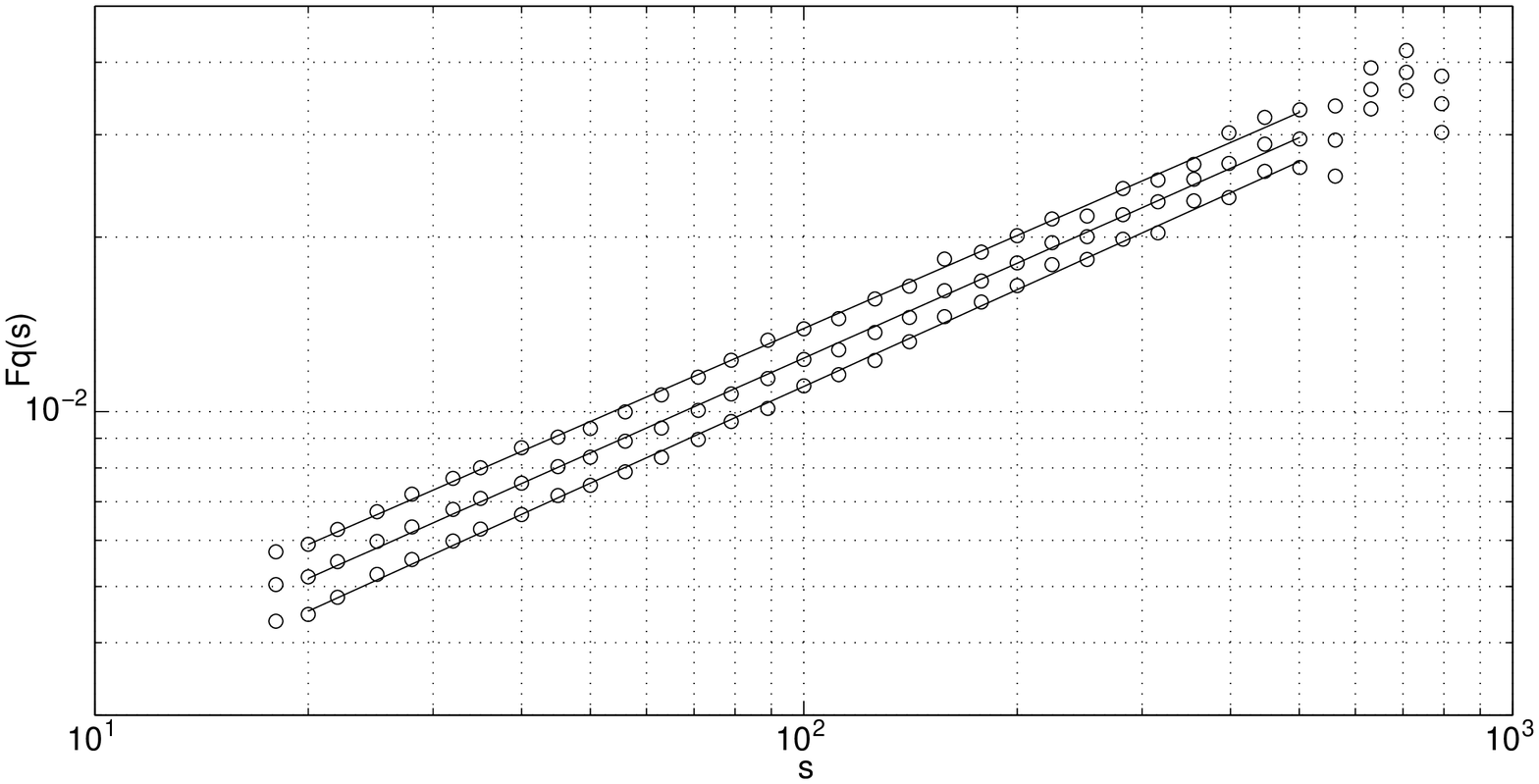} \\
\includegraphics[width=0.5\textwidth,height=0.25\textheight]{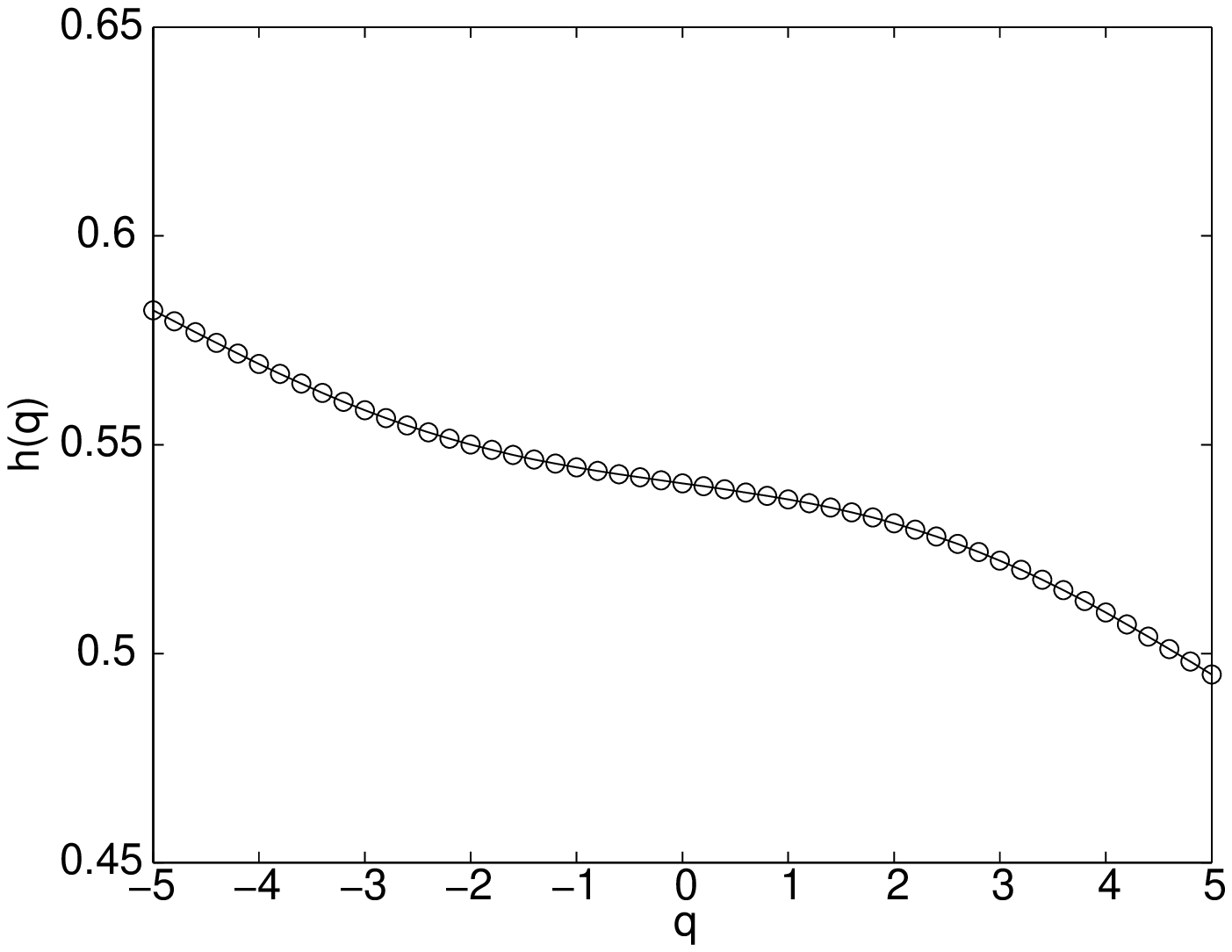} \\
\end{array} $
\end{center}
\begin{picture}(0,0)
   \put(-125,340){\bfseries (a)}
   \put(-125,165){\bfseries (b)}
\end{picture}
\caption{
MFDFA of the exchange rate between the US Dollar and the Euro. ({\bf a}) $q$--order flutuations
for $q$ 2, 0 and -2 (from top to bottom). ({\bf b}) The generalized Hurst exponents
obtained from an application of the MFDFA algorithm.}
\label{fig:hqEURUSDdfa}
\end{figure}

\begin{figure}[t!]
\begin{center}$
\begin{array}{c}
\includegraphics[width=0.5\textwidth]{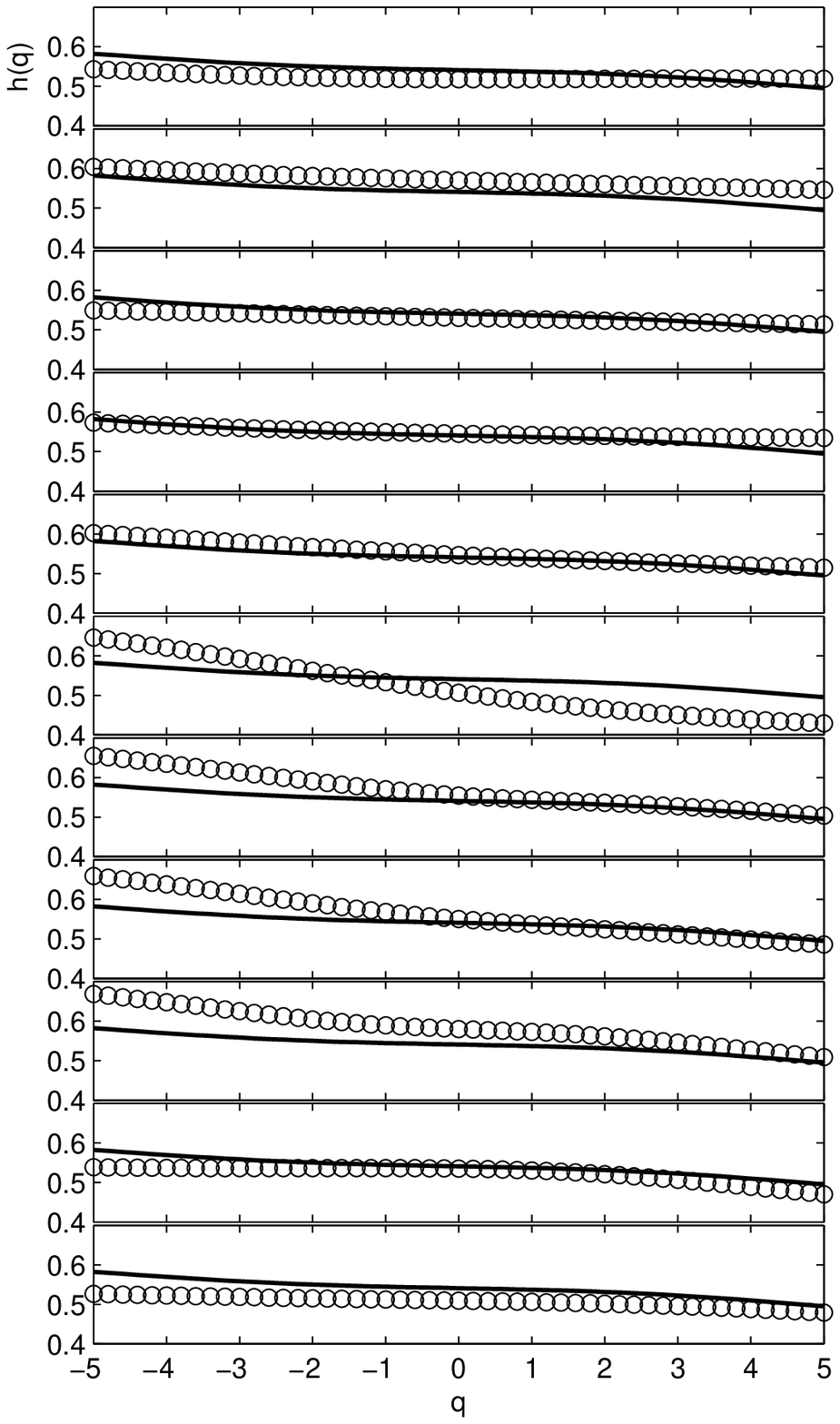}\\
\end{array} $
\end{center}
\begin{picture}(0,0)
   \put(50,430){\bfseries 1999--2002}
   \put(50,395){\bfseries 2000--2003}
      \put(50,360){\bfseries 2001--2004}
   \put(50,325){\bfseries 2002--2005}
   \put(50,290){\bfseries 2003--2006}
   \put(50,255){\bfseries 2004--2007}
   \put(50,220){\bfseries 2005--2008}
   \put(50,185){\bfseries 2006--2009}
   \put(50,153){\bfseries 2007--2010}
   \put(50,115){\bfseries 2008--2011}
   \put(50,80){\bfseries 2009--2012}
\end{picture}
\caption{
Generalized Hurst exponents (open circles) for periods of four years of exchange rates between the US Dollar and the Euro. The
first period starts with the first data available in 1999 (top panel) and the last period starts with the first available data
in 2009. The solid line is the result of the MFDFA method on the complete time series from
1999 to 2012. }
\label{fig:hqEURUSD11}
\end{figure}

\section{Application to the exchange rate between the US Dollar and the Euro
\label{s:Application}}

As mentioned before there is a twofold interest in the study of short time series: ($i$)
many series of interest are short and ($ii$) the dynamics of longer series may change with time requiring the
analysis of shorter pieces of the long series to get an insight into this process. Both aspects are relevant in the
case of exchange rates  because some important currencies are either
relatively new or its exchange to other currency have been subject to new policies in the near
past. The former is the case
for the Euro which was born at the beginning of 1999. The multifractality of the US Dollar to Euro exchange rate has
been studied in \cite{MFDFAUSD} with a different emphasis than here. Other exchange rates involving asian currencies
 have been studied in \cite{MFDFAUSDRial,MFDFAasia}. In particular \cite{MFDFAasia} separetes
 the already short series in two ranges in order to study the effect of the
 Asian currency crisis on the fractal behavior of different Asian exchange rates.

Other more general works which do not only analyze exchange rates, but also other
financial records to study the statistics of return intervals between events above a certain threshold in the context of multifractal models are presented in \cite{bbprl,bbpre78,bbpre80,ltb}.

 \begin{table}[t!]
\caption{\label{t:hq} Values of the mean and standard variation for
$h(q)$ in each period shown in Figure \ref{fig:hqEURUSD11} as well as the corresponding values of
$\Delta_{\rm max} \equiv |\max\{h(q)\}-\left<h(q)\right>|/\left<h(q)\right>$
\vspace{2mm}}
 \begin{ruledtabular}
 \begin{tabular}{|c|c|c|c|}
Period & $\left<h(q)\right>$ & $\sigma(h(q))$ & $\Delta_{\rm max}$ \\
\hline
1999--2002 & 0.522 & 0.007 & 0.039 \\ 
\hline
2000--2003 & 0.572 & 0.017 & 0.057 \\ 
\hline
2001--2004 & 0.531 & 0.011 & 0.034 \\ 
\hline
2002--2005 & 0.548 & 0.012 & 0.045 \\ 
\hline
2003--2006 & 0.551 & 0.026 & 0.092 \\ 
\hline
2004--2007 & 0.519 & 0.069 & 0.244 \\ 
\hline
2005--2008 & 0.566 & 0.045 & 0.156 \\ 
\hline
2006--2009 & 0.560 & 0.052 & 0.176 \\ 
\hline
2007--2010 & 0.584 & 0.043 & 0.144 \\ 
\hline
2008--2011 & 0.522 & 0.020 & 0.100 \\ 
\hline
2009--2012 & 0.507 & 0.013 & 0.055 \\ 
 \end{tabular}
 \end{ruledtabular}
 \end{table}

\subsection{The time series}

The daily exchange rate between the US Dollar
and the Euro is analyzed with the MFDFA method.  The data, shown in the upper
panel of Figure  \ref{fig:hqEURUSD},
has been obtained from the web page of the Board of Governors of the Federal Reserve System (www.federalreserve.gov). It contains 3420
data entries from January 4th, 1999  to August 3rd, 2012. There are approximately 250
entries each year. 
The analysis has been carried out not in the daily exchange rate $r_i$ but on the logarithmic 
differences of the rate in consecutive days $R_i = \ln(r_{i+1})-\ln(r_i)$.
This variable, shown in the lower panel of Figure  \ref{fig:hqEURUSD},
has been chosen in order to be able to compare directly with the results from
\cite{MFDFAUSDRial, MFDFAUSD,MFDFAasia}.

\subsection{Analysis of the full exchange rate time series}

The full time series has been analyzed using the MFDFA method. It has been found that equation (\ref{eq:defhq}) is fulfilled for all 
the range in box sizes $s$, starting from $s=10$, for  $q$ values within -5 and 5. An example
for three $q$ values is shown in the upper panel of Figure \ref{fig:hqEURUSDdfa}.
The fit to extract $h(q)$ has been performed from $s_{\rm min}=20$ to the maximum
available box size $s$. 

The result of the MFDFA is presented in the lower panel of   Figure  \ref{fig:hqEURUSDdfa}.  
The observed behavior is similar to that
observed in the two lower panels of Figure \ref{fig:hqHurst} for the case of monofractal signals with Hurst exponent $H=0.5$ and
$H=0.75$. The region of small $|q|$ has a linear behavior with a very small slope while at higher/lower values of $q$, the generalized Hurst exponent $h(q)$ increases/decreases with a higher slope. The difference between $h(q)$ at $q=0$ and $q=\pm5$ is less than 10\% while the
difference w.r.t $q=\pm2$ is less than 2\%.

  Based on this behavior and the similitudes with the monofractal synthetic series, it is tempting to advance the hypothesis that the series exhibits a monofractal behavior with long-range correlations and a Hurst exponent  around  0.54;
i.e., close to white noise. On the other hand there are many studies with longer financial time series, including exchange rates  (e.g.,  \cite{bbprl,bbpre78,bbpre80})
 which point to a multifractal behavior, and the lower panel of   Figure  \ref{fig:hqEURUSDdfa}
 does not show a constant $h(q)$ so one could also argue the  multifractal scenario. In this
 case, and for this short time series, the multifractality if present would be weak.

\subsection{Analysis of shorter sections of the time series}

From the analysis of monofractal synthetic signals with $H=0.5$ and $H=0.75$, it
was concluded that series
of lengths as short as $2^{10}$ could be analyzed using MFDFA with a precision of some
5\% at the largest values of $|q|<5$, and even better precision for $-3\leq q \le3$. 

For the case of the exchange rates a length of $2^{10}$ corresponds to four years and
a few days of  data.
 So the MFDFA has been applied to segments of $2^{10}$ data
points, where each segment started with the first available data in each year from
1999 to 2009. The last period starting in 2009 and ending in August 3rd, 2012
 has 902 data points. 

It has been found that it was possible to apply the MFDFA method to these shorter time
series and that in each period a behavior as expected from equation (\ref{eq:defhq}) was found.
The generalized Hurst exponents found for each period are shown in
Figure \ref{fig:hqEURUSD11}. 

To quantify somehow the behavior in each period three quantities have been computed:
the mean value of each generalized Hurst exponent over all the $q$, the corresponding
standard deviation and the relative maximum difference between $h(q)$ and the mean.
(In each set $\{h(q)\}$ there were 51
different values of $q$ going from $q=-5$ to $q=5$ in steps of 0.2.) The results are shown in
Table \ref{t:hq}.

The first few periods present a behavior consistent
with a monofractal signal  with a slightly different Hurst exponent. Given the uncertainties expected from the application of MFDFA to short series
is difficult to decide if the differences between these first few periods are due to the shortness of the time series -- which is the likely explanation -- or if they reflect some
deeper dynamics -- which would be very interesting.

But for the period 2004--2007, and in a lesser fashion for those periods surrounding it, the
size of the standard deviation of $h(q)$ for the period or the biggest relative difference between 
the values of $h(q$) and the corresponding mean are simply to big to be explained with the expectations
from monofractal signals. Indeed, for this case the behavior is closer to a Log-Normal expectation.

Note that consecutive periods have a 75\% overlap so their behavior is correlated. This means that the behavior
observed in the period 2004-2007 could have started some time earlier  and lasted for a few years.
 The last  period shows a return to
a monofractal compatible  behavior, quite close to a white noise; i.e. without long-range correlations.  

It is interesting to notice that around 2003 the Euro started to show signals of being a strong currency and also to notice that  the current crisis with the Euro currency started around 2008.
The results found here seem to indicate that the stronger phase of the Euro shows a multifractal
behavior while the weaker phases are closer to a monofractal behavior with a Hurst exponent
just above 0.5. 

A full analysis of these tantalizing indications is
outside the scope of this analysis. For the purposes of this article it is important to notice that the analysis of short time series can
be successfully carried out using the MFDFA technique and that it seems that changes in the fractal behavior of exchange rate
time series can be observed. 

\section{Conclusions\label{s:Conclusions}}

In conclusion, the performance of the MFDFA method has been studied for several
mono- and multifractal models as a function of decreasing length. For all models,
and all lengths, a region in $q$ has been found where the agreement of the simulation
and the theoretical predictions for the
generalized Hurst exponent is of few percent. Outside these regions not only the
agreement is worst, but also the results could led to a wrong assignment of a
multifractal behavior for a monofractal signal or a reduced multifractality
for a multifractal signal.

The results found in this study have been applied to the daily exchange rate
between the USD and the Euro. It has been found that the result of the analysis of 
the series
spanning the 12 years
of existence of the Euro is compatible both, with a monofractal behavior close to white noise and
with a weak multifractal behavior.
Furthermore the analysis of 4 year periods seems to indicate that sometime before 2004
the dynamics of the exchange rate changed from either ($a$) a mono- to a multifractal behavior and that
after some years the dynamics have changed back to a monofractal behavior or ($b$) the multifractal behavior changed from weak to strong to weak in the mentioned periods. 

These results show that with due care, the analysis of short time series is possible with 
MFDFA and that the analysis of short periods of longer time series could help to
discover a change of dynamics in the system under study.

\begin{acknowledgments}
This work has been partially supported by Conacyt Mexico, and by the project LK11209 from the M\v SMT of the
Czech Republic
\end{acknowledgments}

\providecommand{\noopsort}[1]{}\providecommand{\singleletter}[1]{#1}%

\end{document}